\documentclass[11pt]{article}

\usepackage[T1]{fontenc}
\usepackage[cp1251]{inputenc}
\usepackage{textcomp}
\usepackage[centertags]{amsmath}
\usepackage[mediummath]{nccmath}
\usepackage{amsfonts}
\usepackage{amssymb}
\usepackage[hypertex,driverfallback=dvipdfm]{hyperref}
\usepackage{graphicx}
\usepackage[numbers,sort&compress]{natbib}

\usepackage{paperinitial}

\paperinitialization{15mm}{15mm}{15mm}{15mm}{2pt}{10pt}

\DeclareMathOperator{\sgn}{sgn}
\DeclareMathOperator{\Texp}{Texp}

\newcommand{\lan}{\langle}
\newcommand{\ran}{\rangle}

\newcommand{\e}{\varepsilon}
\newcommand{\vf}{\varphi}
\newcommand{\vk}{\varkappa}
\newcommand{\s}{\sigma}

\newcommand{\al}{\alpha}
\newcommand{\be}{\beta}
\newcommand{\ga}{\gamma}

\newcommand{\de}{\delta}
\newcommand{\De}{\Delta}

\newcommand{\la}{\lambda}

\newcommand{\spx}{\mathbf{x}}

\newcommand{\spp}{\mathbf{p}}
\newcommand{\spk}{\mathbf{k}}
\newcommand{\spe}{\mathbf{e}}
\newcommand{\spA}{\mathbf{A}}
\newcommand{\spR}{\mathbf{R}}
\newcommand{\spP}{\mathbf{P}}

\begin{document}
\allowdisplaybreaks[4]

\title{\Large\textbf{Photoexcitation of planar Wannier excitons by twisted photons}}

\date{}

\author{P.O. Kazinski${}^{1)}$\thanks{E-mail: \texttt{kpo@phys.tsu.ru}},\; V.A. Ryakin${}^{1)}$\thanks{E-mail: \texttt{vlad.r.a.phys@yandex.ru}}\\[0.5em]
{\normalsize ${}^{1)}$ Physics Faculty, Tomsk State University, Tomsk 634050, Russia}
}

\maketitle

\begin{abstract}

Photoexcitation of planar Wannier excitons by twisted photons in thin semiconductor films is investigated. The explicit general formulas for transition probabilities between exciton states are derived. The selection rules for the projection of the total angular momentum are obtained. As examples, the Coulomb and Rytova-Keldysh electron-hole interaction potentials are considered. The use of planar excitons as a pure on-chip source of twisted photons is discussed. The formulas obtained also describe photoexcitation of states of electrons and holes bound to charged impurities in planar semiconductors.

\end{abstract}

\section{Introduction}

When a low energy electron and a low energy hole are confined to a semiconducting thin film or a monoatomic layer, they form a planar exciton until they annihilate \cite{Frenkel31,Wannier37,GinzKell73}. Recently, the investigations of properties of such two-dimensional excitons have attracted considerable efforts by many research groups for monolayers of transition metal dichalcogenides and phosphorene \cite{BergMal14,ZWChMR14,Chernikov14,WuQuMac15,Malic18,WangRMP18,Belov19,Molas19,ZhenZhang20,Stepanov21,AnanJoJar21,GrBhSeHa22,Pattan22,QRTKrmp22}. These materials provide the examples of structures with the direct bandgap and the relatively small permittivity entering into the electron-hole interaction potential. As a result, in these structures, the planar excitons can easily be created by electromagnetic waves, they possess large excitation energies and rather long lifetimes \cite{ZhenZhang20,WangRMP18}. The study of properties of these excitons encounters many technical problems. In particular, the exact form of the effective electron-hole interaction potential is unknown and various models describing this interaction are put forward in the literature \cite{GinzKell73,Keldysh79,Stepanov21,Molas19,WangRMP18,Malic18,WuQuMac15,Chernikov14,ZWChMR14,BergMal14,TrPeVe17}. In the present paper, we propose to employ the so-called twisted photons to probe the form of the electron-hole interaction potential.

The twisted photons are the states of the electromagnetic field with definite projection of the total angular momentum \cite{PadgOAM25,Roadmap16,SerboNew,New19,OAMPM}. They are a relatively new instrument for the investigation of properties of solids. Due to the fact that a twisted photon carries a definite projection of the total angular momentum that can be larger than one, it can be used to excite the rotational degrees of freedom of a material in the way inaccessible for dipole transitions and thereby to investigate quantum dynamics of these degrees of freedom. In particular, the twisted photons can produce the excitons in the states with large projection of the total angular momentum \cite{QRTKrmp22,Pattan22,GrBhSeHa22,KonKruGie19}, for example, the dark excitons \cite{ComComDub17,AnanJoJar21,ZhenZhang20,WangRMP18}. We shall show in the present paper that the transition probabilities of photoexcitation of the exciton that has been already created are independent of the Bloch wave functions of the electron and the hole and are determined by the matrix elements of the operator $r^m$. The knowledge of these matrix elements strongly constrains possible models for the electron-hole interaction potential. Furthermore, the transition energies also depend on the projection of the total angular momentum transferred to the exciton in the case when this potential deviates from the Coulomb one. The same conclusions are valid for photoexcitation probabilities of the electron and hole states bound to a charged impurity. Therefore, the twisted photons can be used to probe the interaction potential in this case as well.

Apart from the study of the electron-hole interaction potential, photoexcitation of an exciton is relevant for engineering pure sources of twisted photons. As it will be clear from our study, the inverse process -- deexcitation of the exciton -- gives rise to the creation of a twisted photon. The two-dimensional semiconductor structures open up the possibilities for elaboration of the compact on-chip sources of photons created by excitons \cite{WCBA14,WenWuYu20,AnanJoJar21}. By using two-dimensional excitons, one can construct the pure on-chip sources of single twisted photons or lasers of twisted photons. Such sources are needed for development of telecommunication \cite{Roadmap16,OAMPM,New19} and quantum cryptography \cite{New18,New19}.

The paper is organized as follows. In Sec. \ref{Gen_Form}, we describe the model of a planar Wannier exciton and provide a general formula for its photoexcitation probability. Section \ref{Amplitude_Photo} is devoted to evaluation of the amplitude of photoexcitation of an exciton by a twisted photon for an arbitrary electron-hole interaction potential. In Sec. \ref{ProbPhotoexExc}, we find the probability of photoexcitation introducing the explicit profiles for the exciton center-of-mass wave packet and for the twisted photon. In Conclusion, we summarize the results. In Appendix \ref{Kinet_Contr_App}, we obtain the explicit expression for the kinetic contribution to the photoexcitation amplitude. In Appendix \ref{AppB_Coulomb_Pot}, we analyze in detail the case of the Coulomb electron-hole interaction potential. A comparison of the photoexcitation probabilities corresponding to the Coulomb and Rytova-Keldysh (RK) interaction potentials is given in Appendix \ref{AppC_Ryt_Keld} .

We use the system of units such that $\hbar=c=1$ and $e^2=4\pi\alpha$, where $\alpha\approx1/137$ is the fine structure constant. However, in the estimates, we will restore the dependence on the velocity of light $c$.


\section{General formulas}\label{Gen_Form}

Let us consider a two-dimensional Wannier exciton \cite{Wannier37} interacting with the quantum electromagnetic field in a thin semiconductor film or in a semiconducting monolayer. It is assumed that the exciton is localized in the $(x,y)$ plane, i.e., the corresponding Bohr radius (see Eq. \eqref{Bohr_radius} below) is much larger than the thickness of the film. The Bloch functions of the electron and the hole in the periodic potential of the crystal have the form
\begin{equation}
	\vf_e(\spp,\spx_1)=e^{i\spp\spx_1}u_{\spp}(\spx_1),\qquad \vf_h(\spp,\spx_2)=e^{i\spp\spx_1}v_{\spp}(\spx_1),
\end{equation}
respectively, where $u_{\spp}(\spx_1)$ and $v_{\spp}(\spx_2)$ are the periodic functions of $\spx_{1,2}$. We do not show the index that enumerates the bands. Hereinafter, the coordinates and the momenta corresponding to an electron and a hole are supposed to be two-dimensional. The exciton state is approximately described by the wave function \cite{Kittel63,Knox63,Elliot63,QRTKrmp22}
\begin{equation}\label{excit_wavefunc}
	\Psi(\spx_1,\spx_2)=u_{\spp_{01}}(\spx_1) v_{\spp_{02}}(\spx_2) F(\spx_1,\spx_2),
\end{equation}
where the quasimomenta $\spp_{01}$ and $\spp_{02}$ correspond to the lower edge of the conduction band and the upper edge of the valence band, respectively. The function $F(\spx_1,\spx_2)$ satisfies the equation
\begin{equation}\label{Wannier_Ham}
	\Big[\frac{\de\spp_1^2}{2m_1} + \frac{\de\spp_2^2}{2m_2} + V(|\spx_1 - \spx_2|)+const\Big]F(\spx_1,\spx_2)=E_WF(\spx_1,\spx_2),
\end{equation}
where $\de\spp_{1,2}=\spp_{1,2}-\spp_{01,02}$. It is assumed in Eq. \eqref{Wannier_Ham} that the dispersion law of the electron possesses the spherical symmetry near the bottom of the conduction band, $\spp_{01}$, and near of the top of the valence band, $\spp_{02}$. Furthermore, $m_1$ and $m_2$ are the effective masses of the electron and the hole. We neglect the exchange and spin contributions assuming that they are small \cite{AmaMar17,HonerPel18,Elliot63,Pattan22,WangRMP18,WuQuMac15}.

Without loss of generality, the constant in \eqref{Wannier_Ham} can be set to zero. Moreover, it is convenient to change the representation such that
\begin{equation}
	\de\spp_{1,2}\rightarrow\spp_{1,2},
\end{equation}
i.e., to perform the shift in the momentum spaces of the electron and the hole. In this case, implying the Coulomb gauge, the Hamiltonian of the system at issue is written as
\begin{equation}\label{Hamiltonian}
	H = H_0 + H^{(1)}_{int} +H^{(2)}_{int},
\end{equation}
where
\begin{equation}
\begin{split}
	H_0 &= \frac{\spp_1^2}{2m_1} + \frac{\spp_2^2}{2m_2} + V(|\spx_1 - \spx_2|) +H_{em},\\
	H_{int}^{(1)} &= -\frac{e}{2m_1}\big(\hat{\spA}_\perp(\spx_1) \spp_1 + \spp_1 \hat{\spA}_\perp(\spx_1)\big) +
	\frac{e}{2m_2}\big(\hat{\spA}_\perp(\spx_2) \spp_2 +
	\spp_2 \hat{\spA}_\perp(\spx_2)\big),\\
	H_{int}^{(2)} &= \frac{e^2}{2m_1}\hat{\spA}^2_\perp(\spx_1) + \frac{e^2}{2m_2}\hat{\spA}^2_\perp(\spx_2).
\end{split}
\end{equation}
The index $\perp$ stands for the components $(x,y)$ of the tree-dimensional vectors, $H_{em}$ is the Hamiltonian of the free electromagnetic field, $\hat{\spA}(\spx) $ is the operator of the quantum electromagnetic field
\begin{equation}
\begin{split}
	\hat{\spA}(\spx) = \sum_{\ga} \Big[\frac{\spe_{(s)}(\spk)}{\sqrt{2k_{0\ga}V}} \hat{a}_{\ga} e^{i\spk\spx} +
	\frac{\spe^*_{(s)}(\spk)}{\sqrt{2k_{0\ga}V}} \hat{a}^{\dag}_{\ga}e^{-i\spk\spx}  \Big],
\end{split}
\end{equation}
where $V$ is the normalization volume and
\begin{equation}
	\sum_\ga\equiv\sum_s\int\frac{Vd\spk}{(2\pi)^3},\qquad \de_{\ga\ga'}\equiv\frac{(2\pi)^3}{V}\de_{ss'}\de(\spk-\spk').
\end{equation}
The creation and annihilation operators of the photons satisfy the standard commutation relations
\begin{equation}
	[\hat{a}_{\ga},\hat{a}_{\ga'}]=[\hat{a}^\dag_{\ga},\hat{a}^\dag_{\ga'}]=0,\qquad [\hat{a}_{\ga},\hat{a}^\dag_{\ga'}]=\de_{\ga\ga'}.
\end{equation}
The polarization vector is chosen in the form
\begin{equation}
	\spe_{(s)}(\spk)=(\cos\phi_k n_3-is\sin\phi_k,\sin\phi_k n_3+is\cos\phi_k,-n_\perp)/\sqrt{2},
\end{equation}
where $s=\pm1$ is the photon helicity and
\begin{equation}
	\spk=k_0\mathbf{n},\qquad \mathbf{n}=(n_\perp\cos\phi_k,n_\perp\sin\phi_k,n_3),\qquad k_0=|\spk|.
\end{equation}

Using the center-of-mass coordinates, we have
\begin{equation}\label{CM}
\begin{aligned}
	\spR &= \frac{m_1\spx_1 + m_2\spx_2}{M},&\qquad \spx &= \spx_1-\spx_2,\\
	\spp_1 &= \frac{m_1}{M}\spP + \spp,&\qquad \spp_2 &= \frac{m_2}{M}\spP - \spp,
\end{aligned}
\end{equation}
where $M=m_1 + m_2$ and $\spP$ is the center-of-mass momentum. Then
\begin{equation}\label{Hamiltonian1}
\begin{split}
	H_0 =\,&H_{ex} + H_{CM}+H_{em},\\
	H^{(1)}_{int} =\,& -\frac{e}{2}\Big \{ \big[\hat{\spA}_\perp(\spR + \frac{m_2}{M}\spx) - \hat{\spA}_\perp(\spR - \frac{m_1}{M}\spx)
	\big]\frac{\spP}{M} + \frac{\spP}{M}\big[\hat{\spA}_\perp(\spR + \frac{m_2}{M}\spx) - \hat{\spA}_\perp(\spR - \frac{m_1}{M}\spx) \big] +\\
	&+ \big[\frac{\hat{\spA}_\perp(\spR + \frac{m_2}{M}\spx)}{m_1} + \frac{\hat{\spA}_\perp(\spR - \frac{m_1}{M}\spx)}{m_2} \big]\spp
	+\spp\big[\frac{\hat{\spA}_\perp(\spR + \frac{m_2}{M}\spx)}{m_1} + \frac{\hat{\spA}_\perp(\spR - \frac{m_1}{M}\spx)}{m_2} \big]\Big\},
\end{split}
\end{equation}
where
\begin{equation}
	H_{ex}=\frac{\spp^2}{2\mu} + V(|\spx|),\qquad H_{CM}=\frac{\spP^2}{2 M},
\end{equation}
and $\mu^{-1}=m_1^{-1}+m_2^{-1}$.

We standardly construct the perturbation theory taking $H_0$ as an unperturbed Hamiltonian. The evolution operator is given by
\begin{equation}
	U_{t_2,t_2}=U^0_{t_2,0}S_{t_2t_1}U^0_{0,t_1},\qquad
    S_{t_2,t_1}=\Texp\Big\{-i\int_{t_1}^{t_2}d\tau\big[H_{int}^{(1)}(\tau)+H_{int}^{(2)}(\tau)\big]\Big\},
\end{equation}
where $U^0_{t_2,t_1}$ is the free evolution operator, and the interaction Hamiltonian is written in the interaction representation constructed by means of $H_0$. Hereafter, we consider the first Born approximation with respect to the electron charge. In this case, we have
\begin{equation}
	S_{t_2,t_1}\approx1-i\int_{t_1}^{t_2}d\tau H_{int}^{(1)}(\tau).
\end{equation}

Let us given the orthonormal set of eigenfunctions of the exciton Hamiltonian $H_{ex}$ with the definite values of the projection of angular momentum onto the $z$ axis
\begin{equation}\label{excit_eigenfunc}
	H_{ex}\psi_{nl}=E_{nl}\psi_{nl},\qquad L_z\psi_{nl}=l\psi_{nl}.
\end{equation}
Suppose that the system is prepared at $t=t_1$ in the state
\begin{equation}
	|in\ran=\sum_\ga e^{-ik_{0\ga}t_1}\vf^{in}_\ga \hat{a}^\dag_\ga|0\ran e^{-i E_{nl}t_1}\psi_{nl}(\spx) \int\frac{d\spP}{2\pi}f(\spP) e^{-i E_{CM}(\spP)t_1+i\spP\spR} u_{0}(\spx_1)v_{0}(\spx_2),
\end{equation}
where $E_{CM}(\spP)=\spP^2/(2M)$, $|0\ran$ is the vacuum state of photons, $\vf^{in}_\ga$ specifies the wave packet of the incident photon at $t=0$, and $f(\spP)$ defines the wave packet of the exciton center of mass at $t=0$. Furthermore,
\begin{equation}\label{norm_conds}
	\sum_\ga|\vf^{in}_\ga|^2=1,\qquad\int d\spP |f(\spP)|^2=1.
\end{equation}
As for the final state, we take it in the form
\begin{equation}
	|out\ran=|0\ran e^{-i E_{n'l'}t_2}\psi_{n'l'}(\spx) \frac{e^{-i E_{CM}(\spP')t_2+i\spP'\spR}}{\sqrt{S}}u_{0}(\spx_1)v_{0}(\spx_2),
\end{equation}
where $S$ is the area of the film that contains a two-dimensional exciton. Therefore, the transition amplitude in the first Born approximation is written as
\begin{equation}
\begin{split}
	-i\lan out|\int_{-\infty}^{\infty}d\tau H_{int}^{(1)}(\tau)|in\ran=\,&-(2\pi)^2i\sum_\ga\int \frac{d\spP}{\sqrt{S}} \de\big(E_{CM}(\spP')+E_{n'l'}-E_{CM}(\spP)-E_{nl}-k_{0\ga}\big)\times\\
	&\times\de\big(\spP'-\spP-\spk_\perp\big) A(\spk,\spP)  \frac{\vf^{in}_\ga f(\spP)}{\sqrt{2k_{0\ga}V}},
\end{split}
\end{equation}
for $t_1\rightarrow-\infty$ and $t_2\rightarrow\infty$, where
\begin{equation}\label{A_ampl}
\begin{split}
	A(\spk,\spP)=&-\frac{e}{2}\int d\spx \psi_{n'l'}^*(\spx)\spe^{(s)}_\perp(\spk) \big[\frac{\spP' + \spP}{M}\big(e^{i\frac{m_2}{M}\spk_\perp\spx} - e^{-i\frac{m_1}{M}\spk_\perp\spx}\big) +\\
	&+ \big(\frac{e^{i\frac{m_2}{M}\spk_\perp\spx}}{m_1} + \frac{e^{-i\frac{m_1}{M}\spk_\perp\spx}}{m_2}\big)\spp +
	\spp\big(\frac{e^{i\frac{m_2}{M}\spk_\perp\spx}}{m_1} + \frac{e^{-i\frac{m_1}{M}\spk_\perp\spx}}{m_2}\big)  \big]\psi_{nl}(\spx),
\end{split}
\end{equation}
and $\spP'=\spP+\spk_\perp$.

In deriving formula \eqref{A_ampl}, we have employed the fact that the exciton wave function, $\psi_{nl}(\spx)\exp(i\spP\spR)$, and the incident photon wave function vary slowly on the scale of a crystal cell. For such functions, $f(\spx)$, we have
\begin{equation}
\begin{gathered}
	\int d\spx u^*_0(\spx)f(\spx) \mathcal{O}u_0(\spx)=\sum_{a} f(\spx_a) \int_{\Omega_a} d\spx
    u^*_0(\spx)\mathcal{O}u_0(\spx)=\lan\mathcal{O}\ran\int d\spx f(\spx),\\ \lan\mathcal{O}\ran:=\int_{\Omega_a}\frac{d\spx}{v}u^*_0(\spx)\mathcal{O}u_0(\spx),
\end{gathered}
\end{equation}
where the sum is taken over all the crystal cells $\Omega_a$ with the volume $v$ and it is assumed that the operator $\mathcal{O}$ is invariant under displacements by the lattice vectors. It is clear that $u_0(\spx)=\vf_e(0,\spx)$, $v_0(\spx)=\vf_h(0,\spx)$. In particular, for the $\mathcal{O}=\spp_1$, we deduce
\begin{equation}
	\lan\spp_1\ran=\spp|_{\spp=0}=0,
\end{equation}
where the following relation is employed \cite{Kittel63,LandLifshST2}
\begin{equation}
	\int d\spx\vf^*_e(\spp,\spx)\dot{x}^i\vf_e(\spp,\spx)=\frac{\partial\e(\spp)}{\partial p_i},
\end{equation}
and $\e(\spp)$ defines the dispersion law of an electron. The analogous formulas are valid for the integrals containing the hole wave functions $v_0(\spx)$. As a result, the amplitude \eqref{A_ampl} does not depend on the explicit expressions of the Bloch functions of the electron and the hole entering into the states $|in\ran$ and $|out\ran$. It is a consequence of the fact that the wave function of the exciton \eqref{excit_wavefunc} involves the periodic parts of the Bloch wave functions taken at the edges of the valence and conduction bands where the average velocities of the electron and the hole vanish.

The inclusive probability of photoexcitation of the two-dimensional exciton,
\begin{equation}
	P_{n'l',nl}=\int\frac{Sd\spP'}{(2\pi)^2} \Big|\lan out|\int_{-\infty}^{\infty}d\tau H_{int}^{(1)}(\tau)|in\ran\Big|^2,
\end{equation}
reads
\begin{equation}\label{prob_photoexcit}
\begin{split}
	P_{n'l',nl}=&\,(2\pi)^2\sum_{\ga_1\ga_2} \int d\spP' d\spP_1
    d\spP_2\de\big(\frac{\spP'^2}{2M}+E_{n'l'}-\frac{\spP_1^2}{2M}-E_{nl}-k_{0\ga_1}\big)\de(\spP'-\spP_1-\spk_{\perp1})\times\\
	&\times \de\big(\frac{\spP'^2}{2M}+E_{n'l'} -\frac{\spP_2^2}{2M}-E_{nl}-k_{0\ga_2}\big) \de(\spP'-\spP_2-\spk_{\perp2})\times\\
	&\times A(\spP_1,\spk_{\perp1})
    A^*(\spP_2,\spk_{\perp2})\frac{\vf^{in}_{\ga_1}\vf^{*in}_{\ga_2}f(\spP_1)f^*(\spP_2)}{2V\sqrt{k_{0\ga_1}k_{0\ga_2}}}.
\end{split}
\end{equation}

\section{Amplitude of photoexcitation of the exciton}\label{Amplitude_Photo}

Let us find the explicit expression for $A(\spk,\spP)$ in the leading order with respect to the coupling constant. At first, we consider
\begin{equation}\label{amp_def}
	M_\la(\spk) := \int d\spx \psi^{*}_{n'l'}(\spx)\spe^{(s)}_\perp(\spk) (\spp e^{i\la\spk_\perp\spx}
    +e^{i\la\spk_\perp\spx}\spp)\psi_{nl}(\spx).
\end{equation}
In the polar coordinates $(r,\phi)$, it follows from \eqref{excit_eigenfunc} that
\begin{equation}\label{Schrod_eqn}
    R''_{nl}+\frac{1}{r}R'_{nl}+\big[2\mu(E_{nl}-V(r))-\frac{l^2}{r^2}\big]R_{nl}=0,\qquad\psi_{nl}=\frac{e^{il\phi}}{\sqrt{2\pi}}R_{nl}(r).
\end{equation}
In order to evaluate the integral \eqref{amp_def}, we employ the Jacobi-Anger expansion,
\begin{equation}\label{Jac_An}
	e^{i\la k_\perp r\cos(\phi_k-\phi)} = \sum_{m=-\infty}^{\infty}i^m J_m(\la k_\perp r)e^{im(\phi_k-\phi)},
\end{equation}
and come to
\begin{equation}\label{p_oper}
	e^{i\la\spk_\perp\spx}\spp\spe^{(s)}_\perp(\spk) = -\frac{i}{2\sqrt{2}}\sum_{m=-\infty}^{\infty}i^m
	J_m(\la k_\perp r)[X_r(\phi)\partial_{r} +
	X_{\phi}(\phi)\frac{1}{r}\partial_{\phi}],
\end{equation}
where
\begin{equation}
\begin{split}
	X_r(\phi) &= (n_3-s) e^{i(m+1)(\phi_k-\phi)} + (n_3+s)e^{i(m-1)(\phi_k-\phi)},\\
	X_{\phi}(\phi) &= i(n_3+s) e^{i(m-1)(\phi_k-\phi)} - i(n_3-s)e^{i(m+1)(\phi_k-\phi)}.
\end{split}
\end{equation}
Further, we apply the operator \eqref{p_oper} to the right in the expression \eqref{amp_def}, whereas the operator standing at first place in parentheses in \eqref{amp_def} is convenient to apply to the left. As a result, we obtain
\begin{equation}
\begin{split}
	M_\la(\spk) =& \frac{i}{4\pi\sqrt{2}}  \sum_{m=-\infty}^{\infty}i^m
	\int dr d\phi rJ_m(\la k_\perp r) \big\{
	e^{il\phi}R_{nl}(r)[X_r(\phi)\partial_{r}+X_{\phi}(\phi)\frac{1}{r}\partial_{\phi}]e^{-il'\phi}R_{n'l'}(r) -\\
	&-e^{-il'\phi}R_{n'l'}(r)[X_r(\phi)\partial_{r}+X_{\phi}(\phi)\frac{1}{r}\partial_{\phi}]e^{il\phi}R_{nl}(r)\big\}.
\end{split}
\end{equation}
Opening brackets, we see that there are only two different integrals over $ r $ and the two ones over $ \phi $:
\begin{equation}\label{ints}
\begin{split}
	A^m_{n'l'nl} &= \int_{0}^{\infty}dr J_m(\la k_\perp r)R_{n'l'}(r)R_{nl}(r),\\
	B^m_{n'l'nl} &= \int_{0}^{\infty} dr rJ_m(\la k_\perp r)R_{n'l'}(r)R'_{nl}(r),\\
	\Phi^{m}_{ll'} &= \frac{1}{2}\int_{0}^{2\pi} d\phi X_r(\phi)e^{i(l-l')\phi},\\
	\Psi^{m}_{ll'} &= \frac{1}{2}\int_{0}^{2\pi} d\phi X_\phi(\phi)e^{i(l-l')\phi}.
\end{split}
\end{equation}
It is clear that the first integral is symmetric under a simultaneous permutation of $ nl $ and $ n'l' $. Employing this property, we obtain
\begin{equation}
	M_\la(\spk) = \frac{i}{2\pi\sqrt{2}}
	\sum_{m=-\infty}^{\infty}i^m[\Phi^{m}_{ll'}(B^m_{nln'l'}-B^m_{n'l'nl}) - i(l+l') \Psi^{m}_{ll'} A^m_{n'l'nl}].
\end{equation}
The integrals over $ \phi $ are readily evaluated
\begin{equation}
\begin{split}
	\Phi^m_{ll'} &= \pi e^{i(l-l')\phi_k}[(n_3+s) \de_{m,l-l' + 1} + (n_3-s) \de_{m,l-l' - 1}],\\
	\Psi^m_{ll'} &= i\pi e^{i(l-l')\phi_k}[(n_3+s) \de_{m,l-l' + 1} - (n_3-s) \de_{m,l-l' - 1}].
\end{split}
\end{equation}
Due to the fact that
\begin{equation}
	A^m_{n'l'nl} = (-1)^m A^{-m}_{n'l'nl},\qquad B^m_{n'l'nl} = (-1)^m B^{-m}_{n'l'nl},
\end{equation}
we have
\begin{equation}\label{Mpl_tmp}
	M_\la(\spk) = \frac{i}{2\sqrt{2}}
	e^{i(l-l')\phi_k}\sum_{m=0}^{\infty}\sum_{\s=\pm1} i^m
	\de_{m,|l-l' + \s|}(n_3+\s s)[B^m_{nln'l'}-B^m_{n'l'nl} +\s (l+l') A^m_{n'l'nl}].
\end{equation}
The sum over $m$ can also be removed, but we will do it later. Note that the expression \eqref{Mpl_tmp} is antisymmetric under simultaneous permutation
\begin{equation}
	(n,l)\leftrightarrow(n',l'), \qquad s\leftrightarrow-s.
\end{equation}

In order to proceed, let us take into account that the wave functions describing the bound states of an exciton are concentrated at
\begin{equation}\label{Bohr_radius}
	r\lesssim \e r_B,\qquad r_B:=1/(\al\mu).
\end{equation}
The transition energies of an exciton are of the order of (see the explicit expression \eqref{spectrum_Coulomb} for the Coulomb potential)
\begin{equation}\label{est_k0}
	k_0\sim\al/(\e^2 r_B).
\end{equation}
Hence, the arguments of the Bessel functions in \eqref{ints} can be estimated as
\begin{equation}\label{est_r}
	\la k_\perp r =\la k_0 n_\perp r\lesssim \al n_\perp/\e\ll1.
\end{equation}
Therefore, the Bessel functions entering into \eqref{ints} can be replaced by
\begin{equation}\label{Bessel_appr}
	J_m(\la k_\perp r)\approx\frac{1}{m!}\Big(\frac{\la k_\perp r}{2}\Big)^m,\qquad m\geqslant0.
\end{equation}
Moreover, keeping the higher order contributions in the approximate expression \eqref{Bessel_appr} is redundant in the first Born approximation.

Making use of the approximate expression \eqref{Bessel_appr}, we deduce
\begin{equation}\label{A_approx}
	A^m_{n'l'nl}\approx\left\{
						\begin{array}{ll}
							\frac{1}{m!}\Big(\frac{\la k_\perp}{2}\Big)^{m}r^{m-1}_{n'l'nl}, & 	\hbox{\text{for $m\neq1$};} \\
							-\frac12\Big(\frac{\la k_\perp}{2}\Big)^3 r^2_{n'l'nl}, & \hbox{\text{for $m=1$}.}
							\end{array}
						\right.
\end{equation}
where
\begin{equation}\label{r_matrix_elem}
	r^{m}_{n'l'nl}:=\int_0^\infty drr^{m+1}R_{n'l'}(r)R_{nl}(r).
\end{equation}
In Eq. \eqref{A_approx}, we have taken into account that $r^0_{n'l'nl}=0$. Furthermore,
\begin{equation}\label{BB_int}
	B^m_{nln'l'}-B^m_{n'l'nl}\approx\frac{1}{m!}\Big(\frac{\la k_\perp}{2}\Big)^m\int_0^\infty dr r^mrW,
\end{equation}
where
\begin{equation}
	W:=R_{nl}R'_{n'l'}-R'_{nl}R_{n'l'}.
\end{equation}
It follows from the Schr\"{o}dinger equation \eqref{Schrod_eqn} that
\begin{equation}
    (rW)'=\Big[2\mu(E_{nl}-E_{n'l'}) r -\frac{l^2-l'^2}{r}\Big]R_{nl}R_{n'l'}.
\end{equation}
Integrating by parts in \eqref{BB_int}, we come to
\begin{equation}\label{B-B}
	B^m_{nln'l'}-B^m_{n'l'nl}=\frac{(\la k_\perp/2)^m}{(m+1)!}\big[2\mu(E_{n'l'}-E_{nl}) r^{m+1}_{n'l'nl} -(l'^2-l^2)r^{m-1}_{n'l'nl}
	\big].
\end{equation}
Hence,
\begin{equation}\label{M_la_appr}
\begin{split}
	M_\la(\spk)=&\,\frac{ie^{-i(l'-l)\phi_k}}{2\sqrt{2}}\sum_{m=0}^\infty\sum_{\s=\pm1}i^m\de_{m,|l'-l-\s|}\frac{n_3 +\s s}{(m+1)!} \Big(\frac{\la k_\perp}{2}\Big)^m\times\\
	&\times\big[2\mu(E_{n'l'}-E_{nl})r^{m+1}_{n'l'nl}  -(\epsilon-\s) m(l+l')r^{m-1}_{n'l'nl}
	\big],
\end{split}
\end{equation}
where $\epsilon:=\sgn(l'-l-\s)$. In the expression for $M_\la(\spk)$, we have neglected the term arising in formula \eqref{A_approx} for $m=1$ since its contribution is much smaller than the contribution of the first term in the square brackets in \eqref{M_la_appr}.

It follows from the estimates \eqref{est_k0}, \eqref{est_r} that the main contribution to $M_\la(\spk)$ for photoexcitation of an exciton by a plane photon comes from the term with $m=0$. Consequently,
\begin{equation}
	M_\la(\spk)\approx\frac{i\mu}{\sqrt{2}}e^{i(l-l')\phi_k} (E_{n'l'}-E_{nl})r^{1}_{n'l'nl}  \sum_{\s=\pm1}\de_{l',l+\s} (n_3+\s s) .
\end{equation}
Then the contribution on the second line of \eqref{A_ampl} is written as
\begin{equation}\label{second_contr_plane}
	-\frac{e}{2m_1}M_{\frac{m_2}{M}}(\spk)-\frac{e}{2m_2}M_{-\frac{m_1}{M}}(\spk)\approx-\frac{ie}{2\sqrt{2}}e^{i(l-l')\phi_k}
    (E_{n'l'}-E_{nl}) r^1_{n'l'nl}\sum_{\s=\pm1}\de_{l',l+\s}(n_3+\s s).
\end{equation}
As expected, the main contribution is determined by the matrix element of the dipole moment operator of an exciton.

In order to describe photoexcitation of an exciton by twisted photons, one ought to integrate the expression $M_\la(\spk)$ with respect to $\phi_k$ with the factor $\exp(i m_\ga\phi_k)$, where $m_\ga$ is the projection of the total angular momentum of the twisted photon (see \eqref{excit_prob}). Then we find from \eqref{M_la_appr} that the term in $M_\la(\spk)$ giving the leading contribution to this integral has the form
\begin{equation}
	M_\la(\spk)\approx \frac{i^{|m_\ga|}\mu}{\sqrt{2}|m_\ga|!} e^{-i m_\ga\phi_k} \Big(\frac{\la k_\perp}{2}\Big)^{|m_\ga|-1}(n_3+\sgn(m_\ga)s) (E_{n'l'}-E_{nl}) r^{|m_\ga|}_{n'l'nl}\de_{l',l+m_\ga},
\end{equation}
for $m_\ga\neq0$ and
\begin{equation}
	M_\la(\spk)\approx -\frac{\mu\la k_\perp}{2\sqrt{2}} n_3 (E_{n'l}-E_{nl}) r^{2}_{n'lnl}\de_{l'l},
\end{equation}
for $m_\ga=0$. As a result, the contribution on the second line of \eqref{A_ampl} becomes
\begin{equation}\label{second_contr_neq}
\begin{split}
	-\frac{e}{2m_1} M_{\frac{m_2}{M}}(\spk)-\frac{e}{2m_2}M_{-\frac{m_1}{M}}(\spk)\approx&\, -\frac{i^{|m_\ga|}e}{2\sqrt{2}|m_\ga|!}
    e^{-i m_\ga\phi_k}\Big[\Big(\frac{m_2}{M}\Big)^{|m_\ga|}-\Big(-\frac{m_1}{M}\Big)^{|m_\ga|} \Big] \Big(\frac{k_\perp}{2}\Big)^{|m_\ga|-1}\times\\
	&\times(n_3+\sgn(m_\ga)s) (E_{n'l'}-E_{nl}) r^{|m_\ga|}_{n'l'nl}\de_{l',l+m_\ga},
\end{split}
\end{equation}
for $m_\ga\neq0$ and
\begin{equation}\label{second_contr_eq}
	-\frac{e}{2m_1} M_{\frac{m_2}{M}}(\spk)-\frac{e}{2m_2}M_{-\frac{m_1}{M}}(\spk)\approx \frac{ek_\perp n_3}{4\sqrt{2}}
    \Big(\frac{m_2}{M}-\frac{m_1}{M} \Big) (E_{n'l}-E_{nl}) r^{2}_{n'lnl}\de_{l'l},
\end{equation}
for $m_\ga=0$. If the electron and hole masses are approximately equal to each other, it is necessary to take into account the higher order contributions of the perturbation theory to the photoexcitation amplitude for even $m_\ga$. In this case, the excitonic transitions caused by twisted photons with even $m_\ga$ are strongly suppressed.

In the next section, we shall show that the contribution standing on the first line of the photoexcitation amplitude \eqref{A_ampl} is smaller than the contribution on the second line of \eqref{A_ampl} provided the dispersion of momenta in the center-of-mass wave packet of an exciton is much smaller than the de Broglie wavelength of an exciton with the kinetic energy equal to the energy of the absorbed photon. This condition holds in the case when the momentum dispersion of the wave packet of the center of mass is smaller than $\al Mc$. The explicit expression for the contribution on the first line of \eqref{A_ampl} is presented in Appendix \ref{Kinet_Contr_App}.

\section{Probability of photoexcitation of an exciton}\label{ProbPhotoexExc}

We assume that the wave packet of the center of mass of an exciton has the form
\begin{equation}
	f(\spP)=C_e e^{-\frac{\spP^2}{4\s_c^2}},\qquad C_e=1/\sqrt{2\pi\s_c^2},
\end{equation}
i.e., the center of mass of an exciton is at rest on average and it is placed at the origin of the system of coordinates. The state of the incident photon is chosen in the form
\begin{equation}\label{gamma_state}
	\vf^{in}_{\ga_1} =C \de_{ss_1} k_{\perp1}^{|m_\ga|} e^{-\frac{[k_{\perp1}^2-(k_\perp^0)^2]^2}{4 \sigma_\perp^4}}
	e^{-\frac{(k_{31}-k_3^0)^2}{4 \sigma_3^2}} e^{im_\ga \phi_{k1}},\qquad \phi_{k1} = \arg(k_{11} + i k_{21}),
\end{equation}
where $C$ is the normalization constant and $m_\ga$ is the projection of the total angular momentum of the twisted photon onto the $z$ axis. For $\s_\perp$ and $\s_3$ tending to zero, this state turns into a Bessel twisted photon \cite{BKL5,SerboNew}. The normalization constant is found from the condition \eqref{norm_conds}, which is written as
\begin{equation}
	C^2\int\frac{Vd\spk}{(2\pi)^3} k_{\perp}^{2|m_\ga|} e^{-\frac{[k_{\perp}^2-(k_\perp^0)^2]^2}{2 \sigma_\perp^4}}
	e^{-\frac{(k_{3}-k_3^0)^2}{2 \sigma_3^2}}=1.
\end{equation}
In particular, it is clear from this condition that the probability of photoexcitation of an exciton \eqref{prob_photoexcit} does not depend on the normalization volume $V$.

In the expression \eqref{gamma_state}, it is assumed that the quantization axis of the projection of the angular momentum of a twisted photon is perpendicular to the semiconductor film and passes through the center of mass of the exciton. Employing the addition theorem for Bessel functions, it is not difficult to deduce the expression for the photoexcitation amplitude of an exciton in the case when the quantization axis is shifted with respect to the exciton center of mass by the vector $\mathbf{b}$ \cite{BogdKaz19,BKLb}. However, if
\begin{equation}
	k_\perp|\mathbf{b}|\ll1,
\end{equation}
this shift can be neglected.

Let us integrate with respect to $\spP'$ in the general formula for photoexcitation probability \eqref{prob_photoexcit} taking into account one of the delta functions expressing the energy conservation law. Then
\begin{equation}\label{ffbar}
	f(\spP_1)f^*(\spP_2)=C_e^2\exp\Big[-\frac{(\spP_1+\De_{12}\spk_\perp/2)^2}{2\s^2_c} -\frac{\De_{12}\spk_\perp^2}{8\s_c^2}\Big],
\end{equation}
where $\De_{12}\spk_\perp:=\spk_{\perp1}-\spk_{\perp2}$. Suppose that
\begin{equation}\label{k_perp_cond}
	k_\perp/\s_c\ll1.
\end{equation}
In this case, the terms in the exponent in \eqref{ffbar} that contain $\De_{12}\spk_{\perp}$ can be neglected. The condition \eqref{k_perp_cond} implies that the exciton center of mass is localized in the coordinate space on the scale much smaller than the transverse wavelength of the twisted photon.

On integrating with respect to $\spP'$ in \eqref{prob_photoexcit}, the delta functions expressing the energy conservation law lead to the relations
\begin{equation}\label{energy_cons_law}
	k_{01}=\De E +\frac{\spP_{1}\spk_{\perp1}}{M} +\frac{\spk_{\perp1}^2}{2M},\qquad k_{02}=\De E +\frac{\spP_{2}\spk_{\perp2}}{M} +\frac{\spk_{\perp2}^2}{2M},
\end{equation}
where $\De E:=E_{n'l'}-E_{nl}$. The second terms on the right hand side of these equalities describe the Doppler effect due to motion of the exciton center of mass, whereas the third terms arises due to the quantum recoil. These two contributions are negligibly small in comparison with $\De E$. The corresponding conditions when this is the case can be cast into the form
\begin{equation}
	\frac{\s_c n_\perp}{Mc}\ll1,\qquad \frac{\De En_\perp^2}{2Mc^2}\ll1,
\end{equation}
where we have restored the dependence on the velocity of light. Consequently, we can take with good accuracy that
\begin{equation}
	k_{01}=\sqrt{k_{31}^2+k^2_{\perp1}}=k_{02}=\sqrt{k_{32}^2+k^2_{\perp2}}=\De E.
\end{equation}

Further, we suppose that the wave packet \eqref{gamma_state} is narrow with respect to $k_\perp$, i.e., the following estimate holds
\begin{equation}
	\frac{\s_\perp c}{\De E n_\perp}\ll1.
\end{equation}
In this case, one can put $k_{\perp1}=k_{\perp2}=k^0_{\perp}$ in all the expressions in \eqref{prob_photoexcit}, except the Gauss exponent defining the profile of the wave packet with respect to $k^2_\perp$.

The energy conservation law \eqref{energy_cons_law} can be regarded as the equation for $k_{31}$ and $k_{32}$. Then we substitute the corresponding expressions into $\vf^{in}_{\ga_1}\vf^{*in}_{\ga_2}$. Demanding that the contributions of the second and third terms standing on the right hand side of Eqs. \eqref{energy_cons_law} to the exponent in $\vf^{in}_{\ga_1}\vf^{*in}_{\ga_2}$ are negligible, we arrive at the condition
\begin{equation}
	\frac{\De E}{2Mc^2}\frac{\s_c}{\s_3}\frac{n_\perp}{n_3}\ll1.
\end{equation}
This condition imposes the restriction on the dispersion of the longitudinal momentum component of the twisted photon and is not fulfilled in the limit $\s_3\rightarrow0$.

Let us estimate the magnitude of the contribution on the first line of \eqref{A_ampl} to the amplitude of photoexcitation of an exciton in comparison with the contribution on the second line of \eqref{A_ampl}. To this end, we can suppose that
\begin{equation}
	\mu\sim M, \qquad |\spp|\sim\sqrt{2M\De E},\qquad |\spP|\sim\s_c.
\end{equation}
Thus, the contribution on the first line of \eqref{A_ampl} can be neglected in comparison with the contribution on the second line of \eqref{A_ampl} provided that
\begin{equation}\label{first_second_compar}
	\s_c\ll\sqrt{2M\De E}.
\end{equation}
A more accurate comparison of the expressions \eqref{second_contr_plane}, \eqref{second_contr_neq}, \eqref{second_contr_eq} with \eqref{N_la_plane}, \eqref{N_la_neq}, \eqref{N_la_eq} leads to the condition
\begin{equation}
	\s_c\ll\al Mc,
\end{equation}
which is equivalent to \eqref{first_second_compar}. If this condition holds, then
\begin{equation}\label{A_approx_1}
	A(\spk,\spP)\approx-\frac{e}{2m_1}M_{\frac{m_2}{M}}(\spk)-\frac{e}{2m_2}M_{-\frac{m_1}{M}}(\spk).
\end{equation}
The approximate expressions for the terms on the right hand side of this expression are given in \eqref{second_contr_plane}, \eqref{second_contr_neq} and \eqref{second_contr_eq}. Notice that the approximate expression \eqref{A_approx_1} is independent of $\spP$.

As a result, supposing that the estimates discussed above are satisfied, we derive the expression for the probability of photoexcitation of an exciton
\begin{equation}\label{excit_prob}
	P_{n'l',nl}\approx\frac{\De E\s_\perp^2}{4\pi \bar{k}_3^2\s_3}e^{-\frac{(\bar{k}_3-k_3^0)^2}{2\s_3^2}} \int_0^{2\pi} d \phi_{k1}
    d\phi_{k2} e^{i m_\ga(\phi_{k1}-\phi_{k2})}A(\spk_1)A^*(\spk_2),
\end{equation}
where $\bar{k}_3:=\sqrt{(\De E)^2-(k^0_\perp)^2}$. In particular, for the normal incidence of the plane-wave photon $n_3=1$, $m_\ga=s$, we have
\begin{equation}\label{P_plane}
	P_{n'l',nl}=\frac{2\pi^2\alpha\De E\s_\perp^2}{\s_3}e^{-\frac{(\bar{k}_3-k_3^0)^2}{2\s_3^2}} (r^1_{n'l'nl})^2 \de_{s,l'-l}.
\end{equation}
As for the probability of photoexcitation of an exciton by a twisted photon with $m_\ga\neq s$, we find
\begin{equation}\label{P_twisted}
	P_{n'l',nl}=\frac{\pi^2\alpha\De E\s_\perp^2}{2  (|m_\ga|!)^2 n_3^2 \s_3}e^{-\frac{(\bar{k}_3-k_3^0)^2}{2\s_3^2}} \Big[\Big(\frac{m_2}{M}\Big)^{|m_\ga|} -\Big(-\frac{m_1}{M}\Big)^{|m_\ga|} \Big]^2 (n_3+\sgn(m_\ga)s)^2\Big(\frac{k_\perp^0}{2}\Big)^{2|m_\ga|-2} (r^{|m_\ga|}_{n'l'nl})^2 \de_{l',l+m_\ga},
\end{equation}
where $n_3=\bar{k}_3/\De E$. For $m_\ga=0$, we obtain
\begin{equation}\label{P_twisted_1}
	P_{n'l',nl}=\frac{\pi^2\alpha\De E\s_\perp^2}{8 \s_3}e^{-\frac{(\bar{k}_3-k_3^0)^2}{2\s_3^2}} \Big(\frac{m_2}{M} -\frac{m_1}{M} \Big)^2 (k_\perp^{0}r^{2}_{n'l'nl})^2 \de_{l'l}.
\end{equation}
As we see, the following selection rule holds
\begin{equation}\label{sel_rule}
	l'=l+m_\ga.
\end{equation}
Furthermore, as it was mentioned above, when the electron and hole masses are approximately the same, the transition probability is suppressed for even $m_\ga$. The explicit expressions for the matrix elements $r^{m}_{n'l'nl}$ for the Coulomb interaction between an electron and a hole are presented in Appendix \ref{AppB_Coulomb_Pot}. The numerical values for several matrix elements $r^{m}_{n'l'nl}$ in the case of the RK interaction potential are given in Appendix \ref{AppC_Ryt_Keld}.

\section{Conclusion}

Let us sum up the results. We have described photoexcitation of a planar Wannier exciton by plane-wave and twisted photons. We have obtained the explicit expressions \eqref{P_plane}, \eqref{P_twisted}, \eqref{P_twisted_1} for transition probabilities in these cases. These transition probabilities do not depend on the form of the Bloch wave functions of an electron and a hole since we have used the standard approximation that the exciton is composed of the electron and the hole located at the lower edge of the conduction band and the upper edge of the valence band, respectively. We have established that the transition probabilities obey the selection rule \eqref{sel_rule} expressing the conservation law of the projection of the angular momentum. Furthermore, we have found that the transition probability is suppressed for even projections of the total angular momentum of incident photons in the case when the electron and hole masses are equal to each other. As examples, we have considered the Coulomb and RK interaction potentials between an electron and a hole. As for the Coulomb potential, we have derived the explicit expression for the matrix elements $r^m_{n'l'nl}$ determining the transition probability. In the case of the RK potential, we have found several such matrix elements numerically and have shown that for certain $n'l'$ and $nl$ they are larger by orders of magnitude than the same quantity for the Coulomb potential.

As is well known, the Coulomb potential possesses an additional symmetry that results in degeneracy of the energy spectrum with respect to the projection of the angular momentum $l$. In contrast to the Coulomb potential, the spectrum of the Schr\"{o}dinger equation with RK potential is nondegenerate with respect to $|l|$. Therefore, the experimental studies of exciton transitions by means of twisted photons allow one to obtain more detailed information on the form of the electron-hole interaction potential in the planar semiconductor. In particular, we have shown that the ratio of the exciton transition probabilities calculated for different electron-hole interaction potentials is specified by the ratio of the matrix elements $r^m_{n'l'nl}$ provided other parameters are the same.

In the present paper, we did not take into account the spins of the electron and the hole in the exciton, nor was the spin-orbit interaction included \cite{Dyakon17,HonerPel18,Pattan22,WangRMP18}. In particular, we did not study the creation of dark excitons by twisted photons. We leave the investigation of the spin effects for a future work. The other direction of possible studies is the processes of radiation of photons by excitons. It is clear from the above considerations that the exciton transitions provide a source of twisted photons with fixed $|l|$. The degeneracy with respect to the sign of the projection of angular momentum can be removed, for example, by applying the external magnetic field \cite{WangRMP18}. In this case, one obtains a pure source of twisted photons. The other way is to combine the semiconductor thin film or the monolayer with a chiral resonator removing the degeneracy with respect to the sign of $l$ and amplifying the desired twisted mode by means of the Purcell effect \cite{BKL5,BorZhel15,KazKor22}. Stimulated radiation from these transitions can be employed for elaboration of exciton based on-chip lasers \cite{WenWuYu20} of twisted photons or sources of single twisted photons. Notice that the lasers of twisted photons have been already constructed \cite{KorFedShar95,Roadmap16} but their cheap on-chip realization remains a challenge. It is also evident that the formulas for transition probabilities and the conclusions of the present paper are also applicable to the case of photoexcitation of the electron or hole states bound to charged impurities in planar semiconductors. The only change in this case is to put the effective mass of an electron or a hole to infinity.

\appendix
\section{Kinetic contribution to the photoexcitation amplitude of an exciton}\label{Kinet_Contr_App}

Consider the contribution to the amplitude of photoexcitation of an exciton standing on the first line of \eqref{A_ampl}. This contribution is related to the motion of an exciton as a whole. It is easy to see that
\begin{equation}\label{ePP}
	\spe_\perp^{(s)}(\spk)\frac{\spP'+\spP}{M}=\frac{1}{\sqrt{2}} \Big[\frac{k_\perp n_3}{M} +\sum_{\s=\pm1}e^{-i\s\phi_k}(n_3+\s s)
    \frac{P_\s}{M} \Big].
\end{equation}
Let
\begin{equation}
	N_\la(\spk):=\int d\spx \psi^{*}_{n'l'}(\spx) e^{i\la\spk_\perp\spx}\psi_{nl}(\spx).
\end{equation}
Then, taking into account the estimates \eqref{est_k0}, \eqref{est_r} and supposing that $l\neq l'$, we come to
\begin{equation}
	N_\la(\spk)\approx e^{i(l-l')\phi_k} \frac{i^{|l-l'|}}{|l-l'|!} \Big(\frac{\la k_\perp}{2}\Big)^{|l-l'|} r^{|l-l'|}_{n'l'nl},
\end{equation}
and, consequently,
\begin{equation}\label{N_la_neq}
	N_{\frac{m_2}{M}}(\spk)-N_{-\frac{m_1}{M}}(\spk)\approx e^{i(l-l')\phi_k} \frac{i^{|l-l'|}}{|l-l'|!}
    \Big[\Big(\frac{m_2}{M}\Big)^{|l-l'|} -\Big(-\frac{m_1}{M}\Big)^{|l-l'|} \Big] \Big(\frac{k_\perp}{2}\Big)^{|l-l'|} r^{|l-l'|}_{n'l'nl}.
\end{equation}
For $l=l'$, we have
\begin{equation}
	N_\la(\spk)\approx -\Big(\frac{\la k_\perp}{2}\Big)^{2} r^{2}_{n'lnl},
\end{equation}
and
\begin{equation}\label{N_la_eq}
	N_{\frac{m_2}{M}}(\spk)-N_{-\frac{m_1}{M}}(\spk)\approx-\frac{m_2-m_1}{M} \Big(\frac{k_\perp}{2}\Big)^{2}r^{2}_{n'lnl}.
\end{equation}
In the case of photoexcitation of an exciton by plane-wave photons, we obtain
\begin{equation}\label{N_la_plane}
	N_{\frac{m_2}{M}}(\spk)-N_{-\frac{m_1}{M}}(\spk)\approx \frac{ik_\perp}{2} \sum_{\s=\pm1} e^{-i\s\phi_k} \de_{l',l+\s} r^{1}_{n'l'nl},
\end{equation}
i.e., the main contribution comes from the dipole transitions. The complete contribution to the amplitude of photoexcitation of an exciton from the expression on the first line of \eqref{A_ampl} is the product of \eqref{ePP} and \eqref{N_la_neq}, \eqref{N_la_eq} or \eqref{N_la_plane}.

\section{Coulomb potential}\label{AppB_Coulomb_Pot}

In this section, we consider in detail the case when the electron-hole interaction is described by the Coulomb potential
\begin{equation}\label{Coulomb_pot}
	V(r) = e^2/4\pi \e r.
\end{equation}
Then the normalized bound states of a planar exciton have the form \cite{GinzKell73,Ralph65,ShinSug66,YGChWCh91}
\begin{equation}\label{exc_WF_C}
	R_{nl}(r) = \frac{\be_n}{(2|l|)!}\sqrt{\frac{(n+|l|)!}{(2n+1)(n-|l|)!}} (\be_nr)^{|l|}e^{-\frac{\be_n r}{2}}
    F(-n+|l|;2|l|+1;\be_n r),
\end{equation}
where
\begin{equation}
	\be_n =\vk a_n,\qquad \vk=\frac{2\mu\al}{\e}=\frac{2}{\e r_B},\qquad a_n= \frac{1}{n+1/2},\qquad n=\overline{0,\infty}.
\end{equation}
The magnetic quantum number takes the values $l=\overline{-n,n}$. The energy spectrum is written as
\begin{equation}\label{spectrum_Coulomb}
	E_n = -\frac{1}{(n+1/2)^2}\frac{\mu\al^2}{2\e^2}=-\frac{\al}{4\e}\vk a_n^2.
\end{equation}
The confluent hypergeometric function in \eqref{exc_WF_C} is reduced to the Laguerre polynomials \cite{GrRy}
\begin{equation}\label{key1}
	F(-n+|l|;2|l|+1;\be_n r) = \frac{(2|l|)! (n-|l|)!}{(n+|l|)!}L^{2|l|}_{n-|l|}(\be_n r).
\end{equation}
Thus
\begin{equation}
	R_{nl}(r):=p_{nl}\be_n(\be_n r)^{|l|} e^{-\frac{\be_n r}{2}}L^{2|l|}_{n-|l|}(\be_n r),
	\qquad p_{nl}:= \sqrt{\frac{(n-|l|)!}{(2n+1)(n+|l|)!}}.
\end{equation}

Now we turn to evaluation of the integrals over $r$ arising in formula \eqref{ints}. Observe that
\begin{equation}
	\partial_r R_{nl}(r) = \frac{|l|}{r}R_{nl}(r) -\frac{\be_n}{2}R_{nl}(r) - \be_n (\be_n r)^{|l|}
	e^{-\frac{\be_n r}{2}} L^{2|l|+1}_{n-|l|-1}(\be_n r).
\end{equation}
Employing the generating function for the Laguerre polynomials
\begin{equation}
	\sum_{n=0}^{\infty} t^n L^\al_n(x) = (1-t)^{-\al -1}e^{-\frac{xt}{1-t}},
\end{equation}
we obtain
\begin{equation}\label{key3}
	R_{nl}(r) = p_{nl}\be_n (\be_n r)^{|l|}  D^t_{nl}\big[(1-t)^{-2|l|-1} e^{-\frac{\be_n r}{2}\frac{1+t}{1-t}}\big],
\end{equation}
where
\begin{equation}
	D^t_{nl} := \frac{1}{(n-|l|)!}\frac{d^{n-|l|}}{dt^{n-|l|}}\Big|_{t=0}.
\end{equation}

Introduce the notation
\begin{equation}\label{Cm}
	C^m_{n'l'nl,\la}(t,t') := \int_{0}^{\infty}dr r^{|l|+|l'|+\la} J_m(k_\perp r)e^{-\ga_{nn'}(t,t')r},
\end{equation}
where
\begin{equation}
	\ga_{nn'}(t,t') =  \frac{\be_n}{2} \frac{1+t}{1-t} + \frac{\be_{n'}}{2} \frac{1+t'}{1-t'}.
\end{equation}
Then the first integral in \eqref{ints} is written as
\begin{equation}\label{Am}
	A^m_{n'l'nl} = p_{nl}p_{n'l'}\be_n^{|l|+1} \be_{n'}^{|l'|+1}
	D^t_{nl}D^{t'}_{n'l'}\big[(1-t)^{-2|l|-1} (1-t')^{-2|l'|-1} C^m_{n'l'nl0}(t,t')\big].
\end{equation}
The second integral in \eqref{ints} takes the form
\begin{equation}\label{Bm}
\begin{split}
	B^m_{nln'l'}-B^m_{n'l'nl} &= ( |l'|-|l|)A^m_{n'l'nl}+\bar{A}^m_{n'l'nl},\\
	\bar{A}^m_{n'l'nl} &=p_{nl}p_{n'l'}\be_n^{|l|+1} \be_{n'}^{|l'|+1}
	D^t_{nl}D^{t'}_{n'l'}\Big[\frac{\tilde{\ga}_{nn'}(t,t')C^m_{n'l'nl1}(t,t')}{(1-t)^{2|l|+1}(1-t')^{2|l'|+1}} \Big],
\end{split}
\end{equation}
where
\begin{equation}
	\tilde{\ga}_{nn'}(t,t') = \frac{\be_n}{2} \frac{1+t}{1-t} - \frac{\be_{n'}}{2} \frac{1+t'}{1-t'}.
\end{equation}
Therefore, the both radial integrals from \eqref{ints} are expressed in terms of the integral \eqref{Cm}, which in turn is reduced to the Gauss hypergeometric function \cite{GrRy}.

The resulting expression is rather cumbersome. However, the contributions proportional to the second and higher powers of the fine structure constant are redundant in the first Born approximation. Therefore, it is possible to simplify drastically the expressions for the integrals \eqref{ints} and, consequently, for the probability of photoexcitation of an exciton \eqref{P_plane}, \eqref{P_twisted}, \eqref{P_twisted_1}. On stretching the variable $r\rightarrow r/\ga_{nn'}(t,t')$ in the integral \eqref{Cm}, we see that the argument of the Bessel function is much smaller than unity in the region where the integral is saturated. Consequently, the approximation \eqref{Bessel_appr} is applicable. In order to find the probability of photoexcitation of an exciton, it is sufficient to evaluate the matrix elements $r^{m}_{n'l'nl} $. The corresponding integral \eqref{r_matrix_elem} boils down to the gamma function. As a result, we come to
\begin{equation}\label{Coulomb_r^m}
\begin{split}
	r^{m}_{n'l'nl} = p_{nl}p_{n'l'}\be_n^{|l|+1} \be_{n'}^{|l'|+1}
	(|l|+|l'| + m+1)! D^t_{nl}D^{t'}_{n'l'}\Big[\frac{(1-t)^{-2|l|-1} (1-t')^{-2|l'|-1}}{\ga_{nn'}^{|l|+|l'| + m+2}(t,t')}\Big].
\end{split}
\end{equation}
This expression can be simplified in some particular cases. For example, the transition from the ground state $n=l=0$ to the state $(n',l')$, $l'\neq0$, is described by the following matrix element
\begin{equation}
	r^{|l'|}_{n'l'00}=\Big(\frac{\e r_B}{4}\Big)^{|l'|}\sqrt{\frac{(n'+|l'|)!}{(n'-|l'|)!}}\frac{|l'|n'^{n'-|l'|-1}}{(1+n')^{n'+|l'|+2}}(1+2n')^{|l'|+3/2}.
\end{equation}
For the transitions from the ground state to the excited states with $ l'=0 $, we obtain
\begin{equation}
	r^2_{n'000} = -\frac{\e^2 r_B^2}{8} \frac{n'^{n'-2}}{(1+n')^{n'+3}}(1+2n')^{7/2}.
\end{equation}

\section{Rytova-Keldysh potential}\label{AppC_Ryt_Keld}

\begin{table}[tp]
\begin{center}
		\begin{tabular}{l|c|c|c|c|c}
			$(n',l')$  &  (1,0) & (1,1) & (2,0) & (2,1) & (2,2)\\
			\hline
			$ E_{n'l'} $, eV    & -0.261  & -0.261  & -0.0941  & -0.0941 & -0.0941\\
			$ \De E_{n'l'} $, eV   & 2.09  & 2.09  & 2.26  & 2.26 & 2.26\\
			$ \vk^m|r^{m}_{n'l'00}| $   & 1.46 & 0.689  &  0.575 & 0.282 & 0.470\\
			\hline
			$ E_{n'l'} $, eV   & -0.169 & -0.215  & -0.0895  & -0.107 & -0.122 \\
			$ \De E_{n'l'} $, eV   & 0.290 & 0.244  & 0.369  & 0.352 & 0.337 \\
			$ \vk^m|r^{m}_{n'l'00}| $  & 9.86 & 2.73 & 3.34 & 0.728 & 10.3\\
			\hline
		\end{tabular}
\caption{{\footnotesize The energies, $E_{n'l'}$, of the exciton states, the transition energies, $\De E_{n'l'}$, from the ground state to the state $(n',l')$, and the values of the matrix elements $r^m_{n'l'nl}$ for different electron-hole interaction potentials. In the upper part of the table, the data for the Coulomb potential are given, whereas in the bottom part of the table, the  data for the RK potential are presented. The energy of the ground state of an exciton in the model with the Coulomb potential is $E_{00}=-2.35$ eV, whereas in the model with the RK potential it is equal to $E_{00}=-0.459$ eV. As follows from \eqref{P_twisted}, \eqref{P_twisted_1}, the index $m=l'-l=l'$ for $l'\neq0$ and $m=2$ for $l'=l=0$.}}	\label{tab:table1}
\end{center}
\end{table}

The field of a point charge in a dielectric plate differs from the Coulomb potential. Imposing the standard boundary conditions on the interfaces of the dielectric plate, it is not difficult to obtain the explicit expression for the electric potential in such a case \cite{Rytova67,Keldysh79}. It is convenient to write the approximate expression for the resulting potential in the form
\begin{equation}\label{KeldPot}
	V_{RK}(r) = - \frac{\al}{r_0}\frac{\pi}{2}\big[ \mathbf{H_0}(\frac{\e r}{r_0}) - Y_{0}(\frac{\e r}{r_0})\big],
\end{equation}
where $ \mathbf{H}_0(x) $ and $ Y_0(x) $ are the Struve and Neumann functions of zeroth order, respectively,
\begin{equation}
	r_0=\e_{pl} d/2,\qquad \e:=(\e_1+\e_2)/2,
\end{equation}
where $d$ is the width of the plate, $\e_{pl}$ and $\e_1$, $\e_2$ are the dielectric permittivity of the plate and of the surrounding media, respectively. For $r\gg r_0/\e$, the potential \eqref{KeldPot} reduces to the Coulomb potential \eqref{Coulomb_pot}. Consequently, as in the case of the Coulomb potential, the Schr\"{o}dinger equation with interaction potential \eqref{KeldPot} possesses an infinite number of bound states at fixed $l$. We enumerate these states by $n$ which by definition $n=n_z+|l|$, where $n_z$ is the number of zeros of the wave function in the region $r>0$, and so $n\geqslant|l|$. Unlike the Coulomb potential, the energy spectrum of the exciton with the electron-hole interaction potential \eqref{KeldPot} is not degenerate with respect to $l$. The energy spectrum is still degenerate only with respect to the sign of $l$.

The probabilities of photoexcitation of an exciton are determined by the matrix elements $r^{m}_{n'l'nl}$. It follows from formulas \eqref{P_plane}, \eqref{P_twisted}, \eqref{P_twisted_1} that the ratio of probabilities of photoexcitation of an exciton found in different models for the electron-hole interaction potential with the same parameters of the incident photons is equal to the ratio of squares of the matrix elements $r^{m}_{n'l'nl}$. The values of these coefficients for transitions from the ground state of an exciton, the transition energies, and the corresponding energies of the states are given in Table \ref{tab:table1}. As an example, we consider the monolayer of MoS${}_2$ on the substrate SiO${}_2$. The parameters of the potential are taken from the paper \cite{WuQuMac15}: $ \e = 2.5 $, $ r_0 = 33.875$ {\AA}. The reduced mass of an electron and a hole in MoS${}_2$ is equal to $ \mu = 0.27 m_e $, where $m_e$ is the electron mass in a vacuum (see, e.g., \cite{LWOTh17}). We see from Table \ref{tab:table1} that for certain transitions the values of the matrix elements $r^{m}_{n'l'00}$ differ by orders of magnitude for the Coulomb and RK potentials.

\end{document}